\documentclass[aps,prd,superscriptaddress,altaffilletter,floatfix,preprintnumbers,amsmath,
amssymb,amsbsy,nofootinbib,nobibnotes,onecolumn,showpacs]{revtex4-1}
\usepackage{graphicx,color,units,bm,subfigure}
\usepackage{graphicx}
\usepackage{ragged2e}
\usepackage{ulem}
\usepackage{subfigure}
\usepackage{caption}
\newcommand{\svnid}[1]{ } 

\begin{document}

\title{StarTrack predictions of the stochastic gravitational-wave background from compact binary mergers}
\date{\today}

\author{C.~P\'{e}rigois}
\affiliation {LAPP,CNRS, 9 Chemin de Bellevue, 74941 Annecy-le-Vieux, France}
\author{C.~Belczynski}
\affiliation {Nicolaus Copernicus Astronomical Center,Bartycka 18,  00-716 Warsaw, Poland }
\author{T.~Bulik}
\affiliation {Astronomical Observatory Warsaw University, Aleje Ujazdowskie 4, 00-478 Warsaw, Poland }
\author{T.~Regimbau}
\email{regimbau@lapp.in2p3.fr}
\affiliation {LAPP,CNRS, 9 Chemin de Bellevue, 74941 Annecy-le-Vieux, France}

\begin{abstract}
We model the gravitational-wave background created by double compact objects 
from isolated binary evolution across cosmic time using the  \textbf{\textit{StarTrack}} binary population code. We include population I/II 
stars as well as metal-free population III stars. Merging and non-merging
double compact object binaries are taken into account.
 In order to model the low frequency signal in the band of the space antenna LISA, we account for the evolution of the redshift and the eccentricity.
We find an energy density of $\Omega_{GW} \sim 1.0 \times 10^{-9}$ at the reference frequency of 25 Hz for population I/II only, making the background detectable at 3 $\sigma$ after about 7 years of observation with the current generation of ground based detectors, such as LIGO, Virgo and Kagra, operating at design sensitivity. 
The contribution from population III is one order of magnitude below the population I/II for the total background, but dominates the residual background, after detected sources have been removed, in 3G detectors.  
It modifies the shape of the spectrum which starts deviating from the usual power law $\Omega_{GW}(f) \sim f^{2/3}$ after $\sim 10$ Hz. The contribution from the population of non merging binaries, on the other hand, is negligible, being orders of magnitude below. Finally, we observe that the eccentricity has no impact in the frequency band of LISA or ground based detectors.

\end{abstract}


\maketitle

\section{ Introduction} 

Gravitational-wave (GW) astronomy started with the first observation of the coalescence of two black holes by Advanced LIGO detectors (aLIGO) \cite{aLIGO} on September 2015 \cite{gw150914}. Two years later, the discovery on August 2017 of the coalescence of two neutron stars \cite{gw170817}, in GWs and light, announced the dawn of multi-messenger astronomy. Advanced LIGO and Advanced Virgo (AdVirgo) \cite{adVirgo} are now detecting signals about every week \cite{O2Catalog} and the frequency will continue to increase until they reach their design sensitivity around 2020-2021. At that time the network will also include the Japanese detector Kagra \cite{Kagra}, already in operation since the summer 2019, and LIGO India (Indigo) \cite{Indigo}, expected to start taking data in 2025.


The sources we are detecting now are loud and close events but represent only the tip of the iceberg. Behind these, a large population of unseen weaker sources at higher redshift is expected to combine to create a gravitational-wave background. The background from the population of compact binary coalescences (CBCs) was investigated by many authors in the past (see ~\cite{2011RAA....11..369R,2011ApJ...739...86Z,2011PhRvD..84h4004R,2011PhRvD..84l4037M,2012PhRvD..85j4024W,2013MNRAS.431..882Z,2015A&A...574A..58K}) and revised after the first detections to account for the high mass observed for black holes and the new rate estimates. The LIGO and Virgo collaborations made predictions of the contribution to the background of binary black holes and binary neutron stars in \cite{Implication1, implication2} with very little assumptions about the model, relying mainly on observations. Other authors investigated the implication of the first detections on the gravitational-wave background, including models of metallicity evolution with redshift and mass distributions \cite{2016arXiv160404288D,2016arXiv160502146N}, and arrived to the same conclusion that the background is likely to be higher than previously expected and has a good chance to be detected within a few years after the design sensitivity of the detectors is reached.

The background from CBCs in the frequency band of terrestrial detectors is not an intrinsic stochastic background, in the sense that the sources do not overlap and can be separated. As the sensitivity of the detectors will improve, more and more sources will be detected, reducing the level of the background dramatically. With third generation detectors such as the planned Einstein Telescope in Europe \cite{ET} or Cosmic Explorer in the US \cite{CE}, we expect to see all of the binary black holes in the universe and a vast majority of binary neutron stars \cite{digging}. In the frequency band of the space antenna LISA (0.1 mHz - 0.1 Hz) \cite{LISA}, where the signals last much longer, we expect however a confusion background \cite{Sesana:2016ljz}.

In this paper, we assume that the mergers we observe are formed through stellar evolution in the field of galaxies, with no dynamical interaction. We use the new population synthesis \textbf{\textit{StarTrack}} to generate realistic populations of merging binaries of all types, binary black holes (BBHs), binary neutron stars (BNSs) and mixed systems with one black hole and one neutron star (BHNSs), and from population I/II and III. Then we calculate the energy density spectrum of the background, accounting for the eccentricity which is expected to play a significant role at low frequencies accessible by the space antenna LISA \cite{LISA}. In addition, for the first time, we include binaries that do not merge within the Hubble time.

In section 1, we summarize the results of the \textbf{\textit{StarTrack}} population synthesis and describe the evolution model, in section 2 we present the spectral properties of the background with and without eccentricity, in section 3 we describe our Monte Carlo procedure used to calculate the energy density of the background, in section 4 we present our results and finally in section 5, we draw our main conclusions. 

\section{The Model} 
\subsection{Population I/II stars}
We use the standard \textbf{\textit{StarTrack}} scenario for modeling the population of stellar compact object binaries. The model has been first presented in \cite{2002ApJ...572..407B} and an improved version of the code is described in \cite{StarTrack}. The current status of the calculations is summarized in \cite{2015ApJ...814...58D} and  \cite{2017arXiv170607053B}. 
The code traces evolution of binaries from the zero age main sequence until formation of compact object binaries and their mergers. We use the standard initial conditions for the initial binaries, as discussed by \cite{2012Sci...337..444S}: \citet{2003ApJ...598.1076K} mass function for the primary, a flat distribution of initial mass ratios, a slightly falling  eccentricity distribution $f(e) \propto e^{-0.42}$, and an initial orbital period  distribution of the form  $f(\log P) \propto (\log P)^{-0.5}$. In the course of the evolution we take into account the mass transfers, and verify their stability. Initial mass transfer are slightly non- conservative. The common envelope episodes are calculated using the formalism developed by \citet{1984ApJ...277..355W}, however we take into account the dependence of the efficiency of the envelope ejection on the structure of the donor. StarTrack distinguishes two different types of treating the common envelope survival depending on the type of the donor: if the donor is on the Hertzsprung gap the CE events either lead to a merger (model B) or allowed to potentially survive (model A)

Formation of the compact object in supernovae explosions includes the fallback, and is described in detail in \cite{2017arXiv170607053B}. The natal kicks in supernovae are described by the \cite{2005MNRAS.360..974H} distribution and are scaled down for black holes with the increasing mass \cite{2017arXiv170607053B}. 
We perform the calculations for a grid of metallicities. Each calculation, labeled $i$, modeled $M_{sim}^i$ mass of stars and led formation of $N^i$ compact objects with the merger time $t^i_k$, and masses $m^i_{1k}$, and $m^i_{2k}$, where $k=1...N^i$.  

The model of the star formation rate history and metallicity evolution  is adopted from \cite{2017ApJ...840...39M}. It is clear that there does not exist a one to one dependence of the metallicity on redshift but that the metallicity distribution evolves with redshift. A detailed study of the metallicity and star formation rate distribution as a function of redshift has recently been shown by \cite{2019MNRAS.488.5300C}.  The model very well described the merger data and is consistent with the constraints on metallicity and star formation evolution. 

With the simulations of binary evolution, using the star formation history and metallicity evolution we calculate the properties of the cosmic compact object binary population. 
In particular, we calculate the rate density in Mpc$^{-3}$ yr$^{-1}$ \cite{2016ApJ...819..108B}:
\begin{equation}
s_i =\frac{1}{M_{sim}} \int_{\Delta Z} SFR(Z)dZ 
\label{rate_density}
\end{equation}
where the integral is the fractional star formation rate (SFR) in the simulated metallicity interval, $M_{sim}$ is the total mass of single and binary stars within the mass range adopted by \textbf{\textit{StarTrack}}  for the initial mass of stars ($0.08-150$ M$_\odot$). The SFR is adopted from extinction corrected model of \cite{Madau:2016jbv}.

\subsection{Population III stars}

For evolution of metal-free (population III) stars we use \cite{1612.01524}. 
In this study the original \textbf{\textit{StarTrack}} code that is used for evolution of population 
I/II stars was extended to allow for evolution of population III stars. 
The initial properties of population III binaries are obtained from dynamical simulations 
of dark matter halos each with several ($\sim 5$) single stars. We employ two very
different models of dark matter halos: large halos (extending to 2000 AU: model FS1) 
and small halos (extending only to 10-20 AU; model FS2). Single star masses are drawn 
from a power-law IMF with slope of $\alpha=0.17$ (top heavy): in range $0.1-140M_\odot$
for model FS1 and $0.1-200M_\odot$ for model FS2 \cite{1211.1889}. 
Dynamical interaction between stars in these dark matter halos allow for the binary 
formation with resultant binary fraction of $\sim 1/3$ \cite{1509.05427}.

In each model initial properties of binaries (initial mass function of primary/more massive 
star in binary, mass ratio, orbital separation and eccentricity) are found to be very different
from each other and also very different for typically assumed initial distributions for Pop I/II 
stars \cite{1207.6397}. These distributions are presented in Figures 5,6,7,8 of \cite{1612.01524}. The evolution of massive stars (progenitors of NSs and BHs) was modified to keep 
stellar radii from excessive expanding based on detailed study of \cite{1402.6672}. This 
limits development of common envelope and formation of merging BH-BH/BH-NS/NS-NS. Additionally, 
wind mass loss was set to zero for population III stars. Besides the above modifications the 
evolution is being treated the same way as for population I/II stars. In this current study we 
employ results of calculations for model FS1.B (where "B" refers to treatment of HG donors in CE 
phase, see Subsection II.a). The BH mass spectrum that corresponds to evolution of single population 
III stars with above prescription is presented in Fig.2 of \cite{1612.01524}. Note that 
maximum BH mass is found to be $\sim 90M_\odot$, consistent with recent LIGO/Virgo discovery of 
$85M_\odot$ BH in the most massive BBH merger discover so far : GW190521 \cite {2009.01190}.

\section{Spectral properties of the background} 

\subsection{Sources that merge within the Hubble time} 
The energy density spectrum of a background of gravitational waves is usually described by the dimensionless quantity \cite{1999PhRvD..59j2001A}:
\begin{equation}
\Omega_\text{GW}(f) = \frac{f}{\rho_c} \frac{d\rho_\text{GW}}{df}\,,
\label{eq:omega1}
\end{equation}
where  $d\rho_\text{GW}$ is the energy density in the frequency
interval $f$ to $f+df$, $\rho _{c} = \frac{3H_0^2c^2}{8\pi G}$ is the critical energy density required to close the Universe, and $H_0$ is the Hubble constant.

For a population of coalescing binaries from all over the Universe and characterized by a set of parameters $\theta$, for instance the component masses, the spins and the initial orbital parameters, we can express it as: 
\begin{equation} 
\Omega_{GW}(f)=\frac{f}{\rho_c H_0} \int d\theta  p(\theta) \int_0^{z_{\rm up}(\theta)} dz \frac{R(z;\theta) \frac{dE_{gw}(f_s;\theta)}{df_s}}{(1+z) E_z(z)}
\label{eq:omega_analyic}
\end{equation}
where we have generalized the usual expression valid for short live sources (see, e.g.~\cite{2011RAA....11..369R,Implication1,implication2}), in order to account for the evolution of the redshift at low frequencies when the sources evolve slowly. In Eq.\ref{eq:omega_analyic}, $p(\theta)$ is the probability distribution of the source parameters, $dE_{gw}/df_s$ is the energy density emitted by a single source, $f_s=f/(1+z)$ is the frequency in the source frame, $R(z;\theta)$ is the rate per unit comoving volume per unit time in the source frame, $z_{\rm up}(\theta)$ is the maximal redshift where a compact binary with parameters $\theta$ can be formed. The factor $(1+z)$ in the denominator converts the rate in the source frame to the detector frame and 
\begin{equation}
E_z(z)=\sqrt{\Omega_M(1+z)^3+\Omega_\Lambda}
\end{equation}
 captures the dependence of the comoving volume on redshift in a $\Lambda$CDM cosmology \cite{1807.06209} with $H_0= 2.183 [s^{-1}]$, $\Omega_M=0.3153$ and $\Omega_\Lambda=0.6847$.
 
The rate $R(z;\theta)$ tracks the cosmic star formation rate, although with a delay between the formation of the massive binary to the time when the source has evolved to the frequency $f_s$.
Accordingly, we write:
\begin{equation}
R(z;\theta)=R_f(z_f(z,\theta))
\end{equation}
The two redshifts $z$ and $z_f$ are connected by the time delay $t_d$ which is the sum of an evolution time $t_b(\theta)$, between the formation of the massive binary at the redshift $z_f$ and the formation of the compact binary at the redshift $z_b$, and the time it takes for the frequency to evolve from the initial frequency $f_i(\theta)$ at the formation of the compact binary, to the emission frequency $f_s$, i.e:
\begin{equation}
\tau_s(f_i(\theta),f_s) =  \frac{5c^5 }{256 (\pi)^{8/3} G^{5/3}}  (\mathcal{M}_c)^{5/3} ((f_i(\theta))^{-8/3} - (f_s)^{-8/3} ) 
\end{equation}
In this expression, $\mathcal{M}_{c} = (m_1m_2)^{3/5} / (m_1+m_2)^{1/5}$ is the chirp mass that depends on the component masses $m_1$ and $m_2$ of the two compact objects. The time delay is also the difference in cosmological lookback times between $z_f$ and $z$:
\begin{equation}
t_d =t_c(z_f) - t_c(z)
\end{equation}
where 
\begin{equation}
t_c(z)= \int_0^{z} \frac{dz'}{H_0(1+z')E_z(z')} dz'
\label{eq:lookback}
\end{equation}

For sources close to the merger, the evolution of the redshift can be neglected and we recover the usual formula (see, e.g.~\cite{2011RAA....11..369R,Implication1,implication2}):
\begin{equation} 
\Omega_{GW}(f)=\frac{f}{\rho_c H_0} \int p(\theta) d\theta  \int_0^{z_{\rm up}(\theta)} dz_m \frac{R_m(z_m;\theta) \frac{dE_{gw}(f_s;\theta)}{df_s}}{(1+z_m) E_z(z_m)}   
\label{eq:omega_analyic1}
\end{equation}
where $z_m$ is the redshift at the time of the merger and $R_m(z_m;\theta)=R_f(z_f(z,\theta))$ is the merger rate. In this case, the delay between $z_m$ and $z_f$ is the sum of the evolution time $t_b(\theta)$ and the merger time between the formation of the compact binary and the merger of the two compact objects.

The spectral energy density spectrum of a single source $dE_{gw}/df_s$, in the case of a circular orbit, is obtained from the relation \cite{phinney}:
\begin{equation}
\frac{1}{4 \pi r^2}\frac{dE_{gw}}{df_s}(f_s) = \frac{\pi c^3}{2G}f_s^2(H_+^2(f_s)+H_\times^2(f_s))
\label{eq:phinney}
\end{equation} 
where $H_+(f_s) = A(f_s) (1+\cos^2\iota) / 2$ and $H_\times(f_s) = A(f_s)\cos \iota $ are the Fourier amplitudes of the two polarization states, $\iota$ is the inclination angle, and $r$ is the proper distance.
Following ~\cite{Implication1,implication2,kowalska2015} we consider the inspiral phase only for BNSs and BHNSs and we use the Newtonian waveforms up to the last stable orbit $f_{\mathrm{ILSO}} = \frac{c^3}{6^{3/2} G \pi M}$, $M=m_1+m_2$ being the total mass, which gives:
\begin{equation}
A(f_s) = \sqrt{\frac{5}{24}} \frac{(G \mathcal{M}_{c})^{5/6}}{\pi^{2/3} c^{3/2}} \frac{1}{r}f_s^{-7/6}
\label{A}
\end{equation}
Replacing in Eq. \ref{eq:phinney}, we obtain ($N,C$ stands for Newtonian and circular): 
\begin{equation}
\frac{dE^{N,C}_{gw}}{df_s} (f_s) = \frac{5 (G \pi)^{2/3} \mathcal{M}_c^{5/3} F_\iota }{12} f_s^{-1/3}
\end{equation}
where $F_{\iota} = (1+\cos^2\iota)^2/4 + \cos^2 \iota $. 

For BBHs, we consider also the merger and ringdown and we use the phenomenological waveforms $A(f)$ of \cite{Ajith2011}, which gives ($P,C$ stands for phenomenological and circular):
\begin{equation}
\frac{dE^{P,C}_{gw}}{df_s} (f_s)=\frac{dE^{N,C}_{gw}}{df_s} (f_s)\left \{
\begin{array}{l l}
 (1+\sum^3_{i=2} \alpha _i\nu^i)^2 & \text{if } f_s<f_{merg}  \\
 f_s  w_m (1+\sum^2_{i=1} \epsilon_i\nu^i)^2 & \text{if }f_{merg}\leq f_s<f_{ring}  \\
 f_s^{1/3} w_r  \mathcal{L}^2(f_s,f_{ring},\sigma) & \text{if } f_{ring}\leq f_s<f_{cut}
 \end{array}
\right . \text{,}\end{equation}
 
with 
\begin{equation}
\nu \equiv (\pi Mf)^{1/3},
\end{equation}

\begin{equation}
\begin{array}{l}
\epsilon_1 = 1.4547\chi - 1.8897,\\
\epsilon_2 = -1.8153\chi + 1.6557,\\
\alpha_2 = -323/224 + 451\eta/168, \\
\alpha_3 = (27/8 -11\eta/6)\chi,
\end{array}
\end{equation}

$\mathcal{L}(f,f_{ring},\sigma)$
is the Lorentz function centered at $f_{ring}$ and with width $\sigma$, $w_m$ and $w_r$ are normalization constants ensuring the continuity between the three phases. 

In the expressions above, 
\begin{equation}
\eta=(m_1m_2)/M^2
\end{equation}
is the symmetric mass ratio and 
\begin{equation}
\chi = \frac{(m_1\vec s_1 + m_2\vec s_2)}{M} \frac{\vec L}{L} 
\end{equation}
is the effective spin, a weighted combination of the projections of the individual spins $\vec s_1$ and $\vec s_2$ on the angular momentum $\vec L$.  

The frequencies at the end of the different phases, inspiral, merger and ringdown, and $\sigma$ ($\mu_k={f_1, f_2,\sigma,f_3}$) are calculated using Eq.2 of \cite{Ajith2011}:
\begin{equation}
\frac {\pi M}{c^3} \mu_k = \mu_k^0 + \sum_{i=1}^3 \sum_{j=0}^{N} x_k^{ij} \eta^i \chi^j 
\end{equation}
where the coefficients $\mu_k^0$ and $x_k^{ij}$ are given in Table I of \cite{Ajith2011}. 

In addition, we account for the dependence on eccentricity, which can play a role before the system has been circularized, for example in the LISA band.
The instantaneous spectrum of gravitational waves from an eccentric binary is given for each harmonic by \cite{2012A&A...541A.120K}:
 \begin{equation}
 \frac{dE^n_{gw}}{df_{s}}(f_{s,n})=  \frac{dE^C_{gw}}{df_{s}} (f_{s}) \frac{g(n,e)}{\Psi(e)} \left(\frac{4}{n^2}\right)^{1/3}
 \end{equation}
with $f_{s,n} =nf_{orb}$. The case n=2 corresponds to the circular orbit.
The function $g(n,e)$ is a sum of Bessel functions:
 \begin{equation}
\begin{array}{c c}
 g(n,e)= \frac{n^4}{32} \left\lbrace\left[J_{n-2}(ne) - 2eJ_{n-1}(ne) + \frac{2}{n}J_n(ne) + 2eJ_{n+1}(ne)-J_{n+2}(ne)\right]^2 \right. \\ 
  \left. +(1-e^2)\left[J_{n-2}(ne) -2eJ_n(ne)+J_{n+2}(ne)\right]^2 + \frac{4}{3n^2}\left[J_n(ne)\right]^2\right\rbrace
\end{array}
\end{equation}
and
\begin{equation}
\Psi(e) = \frac{1+73/74e^2+37/96e^4}{(1-e^2)^{7/2}}
\end{equation}

\subsection{Sources that do not merge within the Hubble time} 
For sources with a lifetime longer than the Hubble time or merging in the future, we assume that the redshift evolves while the frequency stays fixed. In this case, all the compact binaries formed with an initial frequency $f_i$, at any redshift $z_b$ larger than $z$, contribute at redshift $z$ at the same observed frequency $f=(1+z)f_i$, i.e:
\begin{equation} 
\Omega_{GW}(f)=\frac{f}{\rho_c H_0} \int d\theta  p(\theta) \int_0^{z_{\rm up}(\theta)} dz{_b} \int_0^{z_{\rm up}(\theta)} dz \frac{R(z;\theta) \frac{dE_{gw}(f_s;\theta)}{df_s}\delta(f_s-f_i(\theta))}{(1+z) E_z(z)}  
\label{eq:omega_analyic2}
\end{equation}
where we have $R(z;\theta)=R_f(z_f(z_b,\theta))$. In this case the redshift of formation $z_f$ is derived directly from $z_b$, considering that the difference in lookback times (see Eq. \ref{eq:lookback}) between $z_b$ and $z_f$ corresponds $t_b(\theta)$, the time for the massive star binary to evolve into a system of two compact objects:
\begin{equation}
t_b(\theta) = t_c(z_f) - t_c(z_b)
\end{equation}

\section{Simulations} 

In this section we describe the  Monte Carlo procedure we use to estimate the background from a list of sources from the \textbf{\textit{StarTrack}} simulations. With this technique, it becomes easy to add many extra parameters like the spin or the eccentricity, without the burden of having to multiply the number of integrals. Also it allows us to model the evolution of the redshift with the orbital frequency, which is important to extend the calculation to very low frequencies, when the orbital evolution is very slow. And last but not least, we can calculate what we will call the residual background, i.e the background after the sources detected individually have been subtracted.

\subsection{Sources that merge within a Hubble time}

Following \cite{2015A&A...574A..58K} we divide the population of binaries into distinct classes $k$, each corresponding to a source generated by \textbf{\textit{StarTrack}} and characterized by the two component masses $m_1^k$ and $m_2^k$, the redshifts at the time of formation of the massive star $z_f^k$, at the time when the two compact objects are formed $z_b^k$, and at the time of merger $z_m^k$, the eccentricity $e^k$, the metallicity $Z^k$ and the rate density $s_i^k$ defined in Eq. \ref{rate_density}. The sources were generated for a grid of cosmic times separated by $\Delta t=100$ Myr.
In addition, we randomly select the inclination $\iota^k$, the polarization $\psi^k$ and the position in the sky $\Theta^k$ (i.e the declination $\delta^k$ and the right ascension $ra^k$). We assume a uniform distribution of the orientation and an isotropic distribution in the sky \cite{2012PhRvD..86l2001R}. For BBHs, we also draw the unitless spins of the two BHs $\chi_1^k = s_1^k / m_1^k$ and $\chi_2^k = s_2^k / m_2^k$ from a uniform distribution in the range [-1-1].

In order to calculate the individual contributions to the background, we proceed as follow for each class $k$:
\begin{enumerate}
\item we calculate the initial orbital frequency, $f_{orb,b}^k$, at the time of formation of the compact system, and the final orbital frequency, $f_{orb,f}^k$ (the frequency at the last stable orbit for BNSs and BHNSs and the frequency at the end of the ringdown for BBHs).
\item for a grid of frequencies, $f_{orb,j}$, in the range $f_{orb,i}^k - f_{orb,f}^k$, we calculate the corresponding redshift $z(j)$ by solving the equation \cite{2007PhRvD..75d3002R}:
\begin{equation}
\tau_s(z(j),z_b^k) = \frac{5c^5 }{256 (2\pi)^{8/3} G^{5/3}}  (\mathcal{M}_c^k)^{5/3} ((f_{orb,i}^k)^{-8/3} - (f_{orb,j})^{-8/3} )
\label{duration}
\end{equation}
For each harmonic from $n=2$ to 6, we then build a table with 3 columns, the frequency in the source frame $f_s(j)=n f_{orb,j}$, the redshift $z(j)$ and the observed frequency $f(j) =  nf_{orb,j}/(1+z(j))$.

\item for each observed frequency $f$, we can then calculate the contribution of the source to $\Omega_{GW}(f)$ using a discrete version on Eq.\ref{eq:omega_analyic} :
\begin{equation}
\Omega^{n,k}_{GW}(f) =\frac{f}{\rho_c H_0} \frac{s_i^k}{(1+z)E_z(z)} \frac{dz}{dt}(z) \Delta t  \frac{dE^n_{gw}}{df_{s}}(f_{s,n},\theta^k) 
\label{eq:omega_discrete}
\end{equation}
where $z$ and $f_{s,n}$ are obtained from $f$, by interpolating in the table built in the previous step.

\item in order to calculate the residual background we need to remove the sources that can be detected individually. 
For a network of N terrestrial detectors the coherent signal-to-noise ratio (SNR), assuming optimal matched filtering and uncorrelated Gaussian noise in the detectors is given by: 
\begin{equation}
\left( \rho^{k} \right)^2 = \sum_{i=1}^N 4 \int_{f_{i,\min}}^{f_{i,\max}} \frac{\left |F_{+,i}(f,\Theta^k,\psi^k) H^k_{+}(f)
 +F_{\times,i}(f,\Theta^k,\psi^k) H^k_{\times}(f) \right |^2}{S_{n,i}(f)}\ df,
 \label{eq:rho}
\end{equation}
where the index $i$ refers to the detectors, $f_{i,\min}$ and $f_{i,\max}$ are the low and high frequency bounds of their sensitivity band, $F_{+,i}$ and $F_{\times,i}$ are the
antenna response functions to the $+$ and $\times $ polarizations, that depends on the sky position and polarization of the source, and $S_{n,i}(f)$ is the one-sided noise power spectral density (PSD) 
of the $i^{th}$ detector.  We assume that only sources with a SNR below a given threshold $\rho_T=12$ contribute to the residual confusion background.

For LISA, the SNR is given by \cite{Cornish:2018dyw}:
\begin{equation}
   (\rho^{k})^2 = 4 \int_{f_{in}^k}^{f_{in}^k+\Delta f^k}\frac{\left |H^k_{+}(f)
 +H^k_{\times}(f) \right |^2}{S_n(f)}df.
\end{equation}
where $f_{in}^k$ is the frequency of the binary when LISA starts taking data and $f_{in}^k +\Delta f^k$ the frequency after $T_{obs} = 10$ years of the LISA mission.\footnote {In LISA, the frequencies evolve slowly and sources cannot cross the full frequency band over the course of the mission}In order to select the observed frequency $f_{in}^k = (1+z^k) f_{s,in}^k$, we draw uniformly the age of the compact binary from which we calculate the intrinsic frequency $f_{s,in}^k$ and the associated redshift $z^k$ (see Eq.\ref{duration}).

In the expression above, $S_n(f)$ is the effective noise power spectral density including the sky and polarization averaged signal response function of the instrument (see details in \cite{Cornish:2018dyw}). The change in frequency, $\Delta f^k$, is calculated by integrating: 
\begin{equation}
\frac{df}{dt} = \frac{96}{5}\pi^{8/3}(G\mathcal{M}_c^k/c^3)^{5/3}f^{11/3} 
\end{equation}
over the observation time $T_{obs}$.
Following \cite{Cornish:2018dyw}, only the sources with a individual SNR $\rho<$7 contribute to the LISA background.

\item
The number of sources associated to the class $k$ crossing the frequency band of ground-based detectors is simply the total coalescence rate multiplied by the duration of the mission $T_{obs}$:
\begin{equation}
N^k = \left[ \frac{s_i^k}{1+z_m^k}\frac{dV}{dz}(z_m^k)\frac{dz}{dt}(z_m^k)\Delta t \right ]  T_{obs},
\end{equation}

For LISA, because the time the sources spend in band is much larger than the mission lifetime, we instead calculate the number of sources present at any given time: 
\begin{equation}
N_k = \left[ \frac{s_i^k}{1+z^k}\frac{dV}{dz}(z^k)\frac{dz}{dt}(z^k)\Delta t \right ] \tau^k(f_{min},f_{max})
\end{equation}
where $\tau^k(f_{min},f_{max})$ is the time the source spend in the LISA band and $z^k$ is the redshift corresponding to the time when the source is observed, which is drawn uniformly in $\tau(z_b^k)-\tau(z_m^k)$. 
\end{enumerate}

After these steps, we can combine the results of each class.
The energy density parameter is the sum of the contribution from each class and each harmonic:
\begin{equation}
\Omega_{GW}(f) =   \sum_k \sum_{n=2}^6 \Omega^{n,k}_{GW}(f) 
\label{eq:omega}
\end{equation}

and similarly, the total number of sources over the mission lifetime is the sum of the number of sources associated to each class:
\begin{equation}
N = \sum_k N_k 
\end{equation}

\subsection{Sources that do not merge within the Hubble time}

For sources that do not merge within the Hubble time, we assume that the orbital frequency remains constant in time, and only the redshift evolves. We calculate the energy density contribution of each class $k$ for a grid of redshifts between 0 and $z_b^k$, the redshift of formation of the compact object. Then the energy density parameter is obtained following the same procedure as for the population that merge within the Hubble time Eq.\ref{eq:omega}. 

\section{Results} 

\subsection{Total background}
\subsubsection{Sources that merge within the Hubble time}
\begin{figure}
\centering\captionsetup{justification=RaggedRight, name = {Figure},labelfont={bf}}
 \includegraphics[width=15cm]{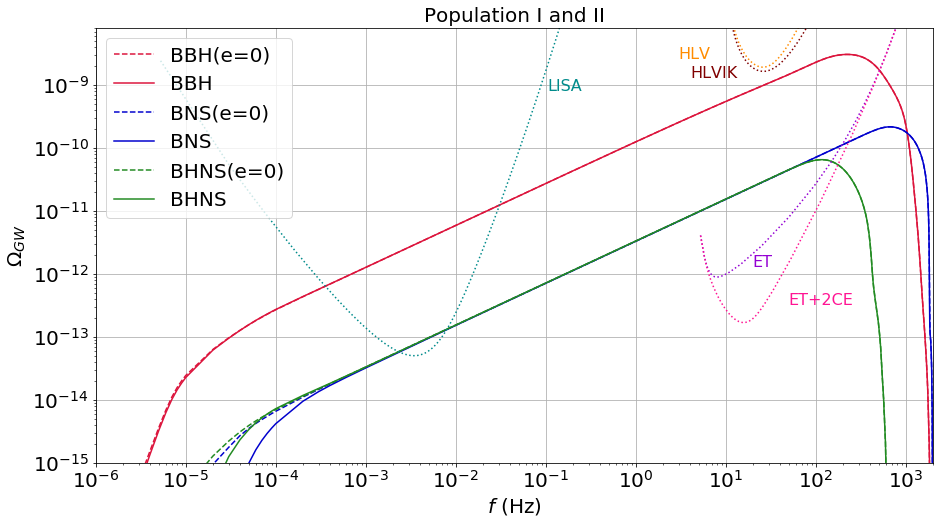}
 \includegraphics[width=15cm]{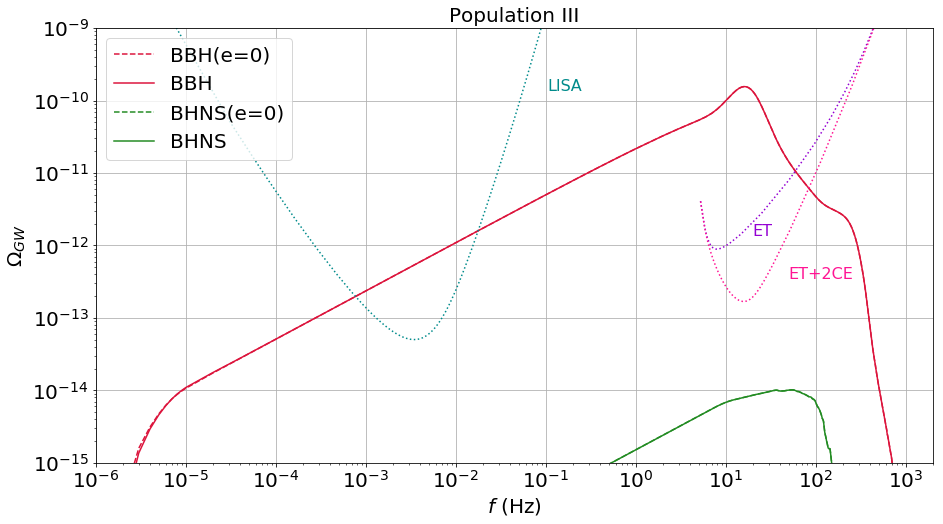}
 \caption{Energy density for the total population of sources that coalesce within the Hubble time. The upper panel is for population I/II stars and the lower panel for population III stars. The three different types of binaries BBHs, BHNSs and BNSs are shown separately in red, green and blue, with a null eccentricity (dashed line) and accounting for harmonics $n=2-5$ (continuous). The dotted lines indicate the Power Integrated curves of the different networks of terrestrial detectors (see text) and LISA.}
\label{fig_poptot}
\end{figure}

The spectra for the different types of binaries (BNSs, BHNSs and BBHs) that coalesce within the Hubble time, including all the sources, are shown in Fig.\ref{fig_poptot} for population I/II stars (top) and for population III stars (bottom). Here we account for the eccentricity (continuous line) and we consider the first four harmonics ($n=2-5$). For comparison we also show the case without eccentricity (dashed line). 
We notice that the eccentricity does not play a significant role, except at the lowest frequencies when the system is still far from being circularized, i.e. below $10^{-3}$ Hz for BNSs, $10^{-4}$ (popI/II) or $10^{-3}$ (popIII) for BHNSs and $10^{-5}$ for BBHs. 

For the three types, and for both populations I/II and III, one can recognize the evolution as $\Omega_{GW}(f) \sim f^{2/3}$, which is characteristic of compact binary models in the inspiral phase. The sharp increase at low frequencies, $\sim 10^{-5}$ Hz for BBHs, $\sim 10^{-4}$  Hz for BHNSs and  $\sim 10^{-3}$ Hz for BNSs, corresponds to frequencies were not all the sources have started to emit GWs (before their initial frequency).
\paragraph*{\bf{Population I/II:}}
It is dominated by BBHs until $f \sim 1$ kHz, where most of them have stopped emitting because they have reached their maximal frequency at the end of the ringdown phase. For BNSs and BHNSs we have considered the inspiral phase only, up to the last stable orbit ;  even without the merger and ringdown regime, because of their smaller mass, BNSs can reach frequencies of $f\approx$2 kHz. Let's notice that \cite{1809.10360} took into account the merger and ringdown in their calculation, assuming the waveforms of \cite{Ajith2011} developped for BBHs can also describe BNSs and BHNSs. The contribution from BNSs in this case extends to $f\approx$10 kHz, but it does not make a difference since detectors are not sensitive to a stochastic background above a few hundred Hz. For BBHs, both studies are consistent with a maximum energy density at 130Hz and 300Hz respectively. 

Our results for BBHs ($\Omega_{GW} (25 \mathrm{ Hz}) = 9.6 \times 10^{-10}$) are in agreement with both the predictions of the LIGO/Virgo collaboration based on the rate and the mass distribution derived from the first two observation runs \cite{2019PhRvD.100f1101A} ($\Omega_{GW}(25 \mathrm{ Hz})= 5.3^{+4.2}_{-2.5} \times 10^{-10}$) and the models of \cite{1607.06818} ($\Omega_{GW}(25 \mathrm{ Hz})$ in $[5 \times 10^{-10}-10^{-8}]$), but they differ from \cite{2019PhRvD.100f1101A} by about one order of magnitude for BNSs, with $\Omega_{GW}(25 \mathrm{ Hz})= 2.8 \times 10^{-11}$ against $\Omega_{GW}(25 \mathrm{ Hz})= 3.6^{+8.4}_{-3.1} \times 10^{-10}$. 
The total of all the contribution, on the other hand, is an order of magnitude below the current upper limit on a stochastic background of $4.8 \times 10^{-8}$ \cite{2019PhRvD.100f1101A}. The reference values of $\Omega_{GW}$ at the most sensitive frequencies for LISA (4 mHz), third generation terrestrial detectors (10 Hz) and second generation terrestrial detectors (25 Hz) are shown in Tables \ref{table-total_Hubble}, for different types of binaries and for Pop I/II and III.

\paragraph*{\bf{Population III:}}
We observe a few bumps, each corresponding to a specific mass range. For example, the first bump at around 10 Hz corresponds to the highest masses in the range ($M>$70M$_\odot$). For comparison, \cite{1603.06921} has derived the contribution from BBHs using the formation model of Kinugawa et al. \cite{1402.6672}, and found a maximum at $f\approx$45Hz. The difference between the redshifted mass distribution of our model ($M_{z,max}^{ST} \approx 140M_\odot$) and Kinugawa et al. ($M_{z,max}^{K} \approx 100M\odot$) explains why we have a maximum at a lower frequency. Other stellar evolution scenarios exist in the literature for population III, for instance \cite{1110.1726} or \cite{osti_115557}, with other redshifted mass distributions favoring lower masses. Using these models, \cite{1604.04288} finds a maximum at 150Hz for the scenario of \cite{1110.1726}  and at 350Hz for the scenario of\cite{osti_115557}. 

The dotted lines in the figure indicate the projected sensitivities, the so-called Power Integrated (PI) curves, for the space antenna LISA and for different terrestrial detector networks: 
\begin{itemize}
\item HLV: Advanced LIGO Hanford (H) and Livingston (L) \cite{aLIGO}, and Advanced Virgo (V) \cite{adVirgo} at design sensitivity. 
\item HLVIK: HLV with in addition LIGO India (I)\cite{Indigo}, whose sensitivity will be similar to the two LIGO detectors, and the Japanese detector Kagra (K) \cite{Kagra}, also at design sensitivity.
\item ET: third generation European detector Einstein Telescope, currently under design study \cite{ET}.
\item ET+2CE: ET and two third generation Cosmic Explorer (CE) \cite{CE}, also under design study.
\end{itemize}
A power-law stochastic background that is tangent to a PI curve is detectable with a signal-to-noise-ratio of 2. For LISA we assume an effective integration time of 5 years (corresponding to the 10 years mission with a duty cycle of about 50\%) and for terrestrial detectors we assume an effective integration time of 1 year following \cite{implication2}.

\begin{table}
\centering\captionsetup{justification=RaggedRight, name = {Table},labelfont={bf}}
\caption{Reference values of the energy density $\Omega_\mathrm{GW}$ for the total population of sources that coalesce within the Hubble time and for the different types of binaries and the sum, at the most sensitive frequencies for LISA (4 mHz), third generation terrestrial detectors (10 Hz) and second generation terrestrial detectors (25 Hz).}

\subtable[Population I/II] {{} \label{table-total_Hubble_12}
\begin{tabular}{|c|ccc|}
 \hline 
$f_{ref}$                     & 4 mHz                    &10 Hz                     & 25 Hz  \\ \hline 
BNS                           & 8.2e-14                    & 1.5e-11              & 2.8e-11               \\ 
BBH                           & 3.2e-12                    & 5.5e-10                & 9.6e-10                \\ 
BH-NS                       & 8.4e-14                    & 1.5e-11              & 2.8e-11               \\ \hline
\textbf{All} 		 & \bf{3.4e-12}       & \bf{5.8e-10}    & \bf{1.0e-9}                \\ \hline
\end{tabular}
}

\subtable[Population III]{{} \label{table-total_Hubble_3}
\begin{tabular}{|c|ccc|}
 \hline   
$f_{ref}$                     & 4 mHz                    &10 Hz                     & 25 Hz  \\ \hline                                      
BNS           & --                    & --                     & --                     \\                                     
BBH           & 5.9e-13         & 9.9e-11           & 8.4e-11           \\
BH-NS       & 3.8e-17         & 6.1e-15         & 9.1e-15         \\ \hline
\textbf{All} & \bf{5.9e-13}    & \bf{9.9e-11}    & \bf{8.4e-11}     \\ \hline
\end{tabular}
}
\label{table-total_Hubble}
\end{table}

\subsubsection{Sources that do not merge within the Hubble time}

\begin{figure}[ht!]
\centering\captionsetup{justification=RaggedRight, name = {Figure},labelfont={bf}}
\includegraphics[width=15cm]{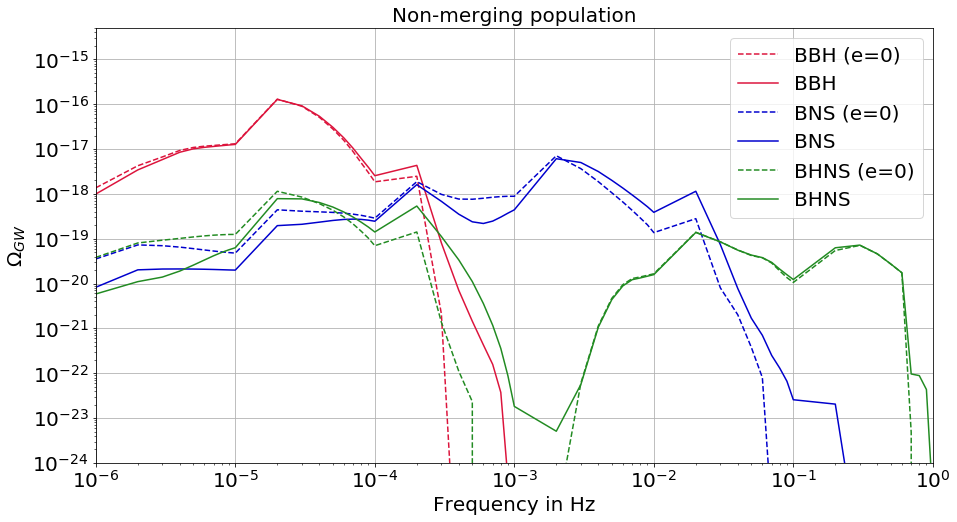}
\caption{Energy density from binaries which do not merge in a Hubble time. The three different types of binaries BBHs, BHNSs and BNSs are shown separately in red, green and blue, with a null eccentricity (dashed line) and accounting for harmonics $n=2-5$ (continuous).}
\label{Nhitime}
\end{figure}

Following the procedure described in section IV, we calculate the contribution of non-merging sources. The orbital evolution of these sources being very slow, they do not contribute at frequencies above 1 Hz. The results are shown in Figure \ref{Nhitime} for the different types of binaries separately (continuous lines). The case of circular orbit (i.e. $e$=0) is also shown for comparison (dashed lines). Because of their higher masses, BBHs contribute at frequencies below 1 mHz, while BNSs contribute to frequencies up to 0.1Hz. For BHNSs, we observe a gap around 1 mHz, separating the sources originating from higher metallicity regions at lower redshift (above 1 mHz) and sources originating from higher redshfit and lower metallicity before 1 mHz.

The energy density is orders of magnitude lower than for the population of sources that merge within the Hubble time, with a maximum of $\Omega_{GW} \sim 10^{-16}$ at 2 $\times$ 10$^{-5}$ Hz against 10$^{-12}$ for the merging sources. 

\subsection{Residual backgrounds}
 
\begin{figure}[]
\centering\captionsetup{justification=RaggedRight, name = {Figure},labelfont={bf}}
\includegraphics[width=15cm]{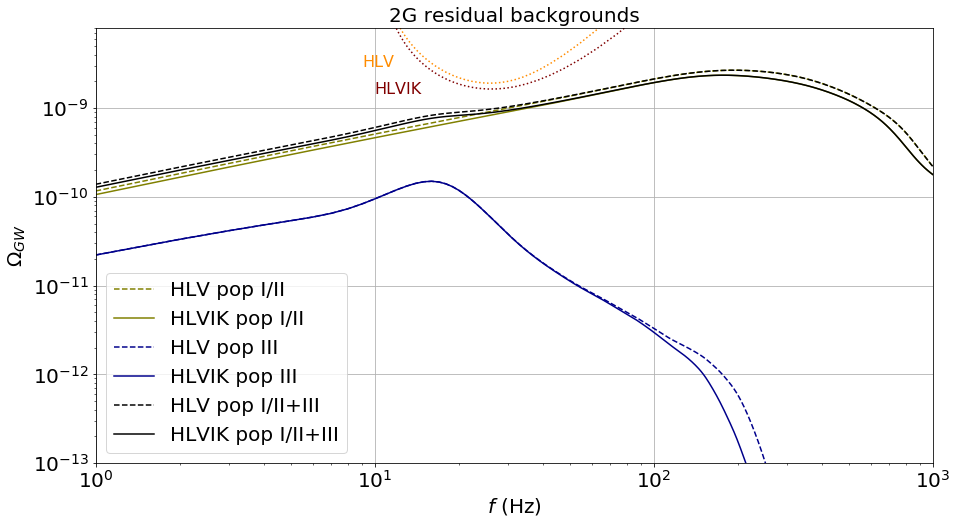}
\includegraphics[width=15cm]{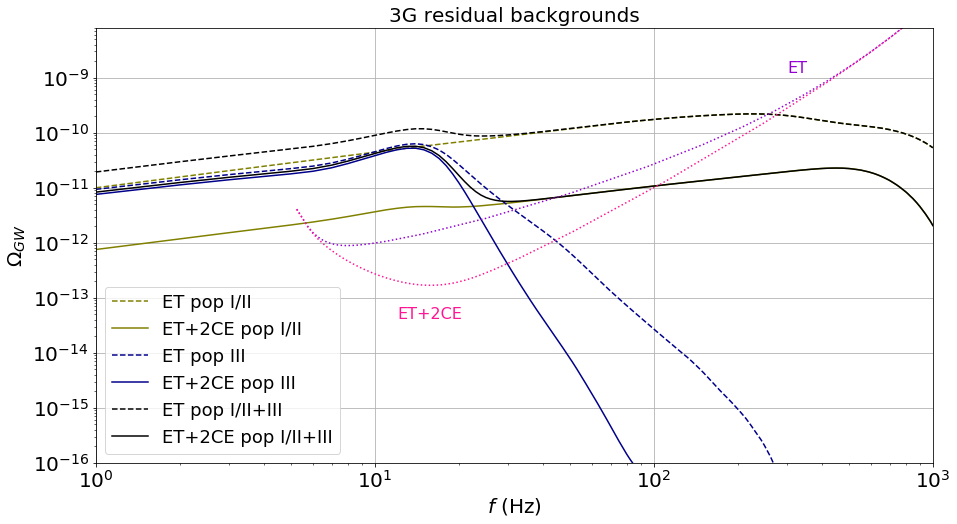}
\caption{Energy density for the residual population I/II and III of sources that coalesce within the Hubble time. The upper panel is for 2G detector networks HLV (dashed lines) and HLVIK (solid lines) and the lower panel for 3G detector networks ET (dashed lines) and ET+2CE (solid lines). The black lines describe the residual energy densities for the total population and the green/blue ones for respectively the population I/II and III.}
   \label{popres}
\end{figure}

As the sensitivity of the detectors will improve in the future, they will be able to detect more sources and therefore decrease the background, assuming one can successfully subtract individual signals from the data \cite{digging,2020arXiv200205365S}. Figure \ref{popres} shows the residual background, i.e the background after individual detections have been removed, for the 2G detector networks HLV and HLVIK (top), and for the 3G detector network ET and ET+2CE (bottom). Here, we have assumed that a source is detected if its signal-to-noise ratio (see Eq. \ref{eq:rho}) is larger than a threshold $\rho_T=12$. For LISA, we obtain that the fraction of detected sources is too small to significantly reduce the background ($\approx$ 5000 sources detected in the 10 years of the mission), in agreement with the predictions of \cite{Sesana:2016ljz}. 

In the frequency band of 3G detectors, the contribution from population III dominates before 40 Hz, reaching a maximum at around 10-20 Hz, and consequently changes the shape of the spectrum that is not a power law anymore. This feature in the shape of the background seems to be characteristic of the Pop. III binaries and can be used as an indicator of their presence. 

 In order to quantify the reduction of the background, we calculate the ratio $r_{\Omega}$ between the energy densities of the residual background and the total background :
\begin{equation}
r_{\Omega}=\frac{\Omega_{GW,res}(f_{ref})}{\Omega_{GW,tot}(f_{ref})},\qquad 
\label{ratio}
\end{equation}
where the reference frequency corresponds to the most sensitive frequency of the network $f_{ref}$. Values of $r_\Omega$ for the total populations I/II and III are shown in Table \ref{resfactors}. 
With second generation detectors, the reduction is small: the residual background is only 0.8-0.9 (HLVIK-HLV) time smaller than the total background and most of the sources removed in this case are BBHs (see below). With the third generation, the reduction is significant, with $r_{\Omega}$ of the order of 0.01-0.1(ET+2CE-ET).

\begin{table}
\centering\captionsetup{justification=RaggedRight, name = {Table},labelfont={bf}}
\caption{Ratio between the energy densities of the residual background and the total background for population I/II and III, quantifying the reduction of the background, for different networks of detectors, evaluated at the most sensitive frequency of the network (25 Hz for second generation and 10 Hz for third generation).}

\begin{tabular}{|c|cc||cc|}
\hline
Network                       & HLV        & HLVIK      & ET         & ET+2CE     \\ \hline
$f_{ref}$ (Hz)              &  25      & 25   & 10  & 10  \\
pop I/II & 0.88      & 0.80      & 0.08      & 0.06      \\
pop III       & 0.80        & 0.79          & 0.46      & 0.39  \\    
\hline
\end{tabular}
\label{resfactors}
\end{table}

\begin{figure}[]
\centering\captionsetup{justification=RaggedRight, name = {Figure},labelfont={bf}}
\includegraphics[width=15cm]{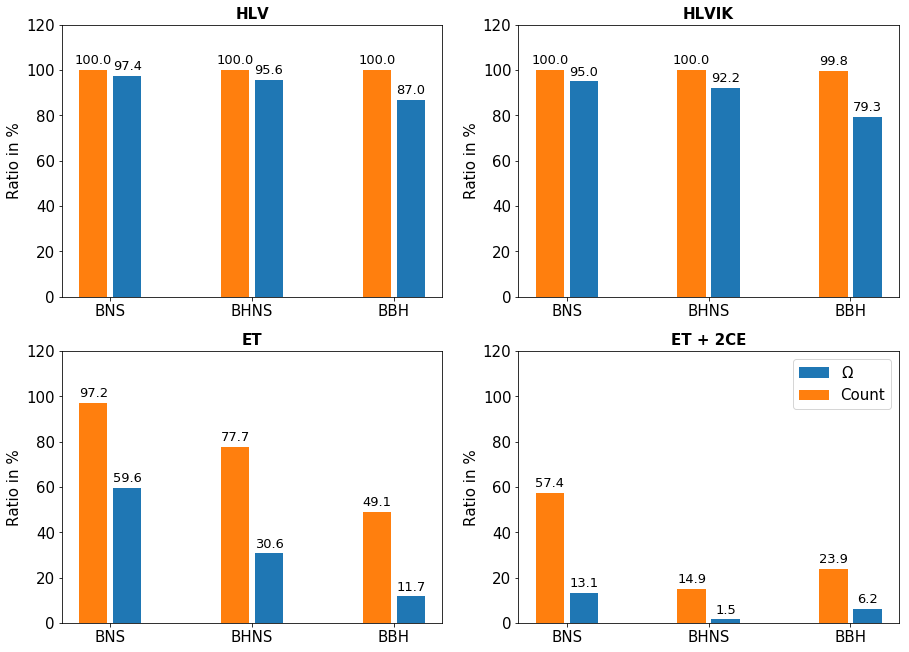}
\caption{Ratios $r_{Count}$ (blue) and $r_{\Omega}$ (orange) for the terrestrial detector residual backgrounds for the total population I/II and III.}
   \label{fig_ratios}
\end{figure}

For comparison, we also calculate the ratio between the number of sources contributing to the residual background and the total number of sources:
\begin{equation}
 r_{Count}=\frac{N_{res}(f_{ref})}{N_{tot}(f_{ref})}.
\label{ratio_Count}
\end{equation}
Figure \ref{fig_ratios} compares the ratios $r_{\Omega}$ (orange bars) and $r_{Count}$ (blue bars) for the three types of binaries BNSs, BBHs and BHNSs, and for the different terrestrial detector networks considered in this study i.e HLV, HLVIK, ET and ET+2CE. Because the sources that are detected at the lowest redshifts are also those whose contribution to $\Omega_{GW}$ is the largest, the ratio $r_{Count}$ is higher than $r_{\Omega}$ for every type of binary and detector network. 

With second generation detectors, only a very small fraction of sources can be resolved with $r_{Count}<0.1\%$ and the reduction of $\Omega_{GW}$ is at most 20.7-8.8\% for BBHs-BHNSs (because of their higher masses), and for HLVIK. We notice that adding the two detectors Indigo et Kagra does not decrease significantly the residual background. Detectors of the third generation can resolve a larger fraction of the sources, leaving 11.7\% of BBHs, 30.6\% of BHNSs and 59.6\% of BNSs in the case of ET, and 6.2\% of BBHs, 1.5\% of BHNSs and 13.1\% of BNSs in the case of ET+2CE. This corresponds to $r_{\Omega}$ of 49.1\% for BBHs, 92.2\% for BHNSs and 23.9\% for BNSs in the case of ET, and 23.9\% for BBHs, 14.9\% for BHNSs and 57.4\% for BNSs in the case of ET+2CE.

\begin{figure}[]
\centering\captionsetup{justification=RaggedRight, name = {Figure},labelfont={bf}}
\includegraphics[width=7.5cm]{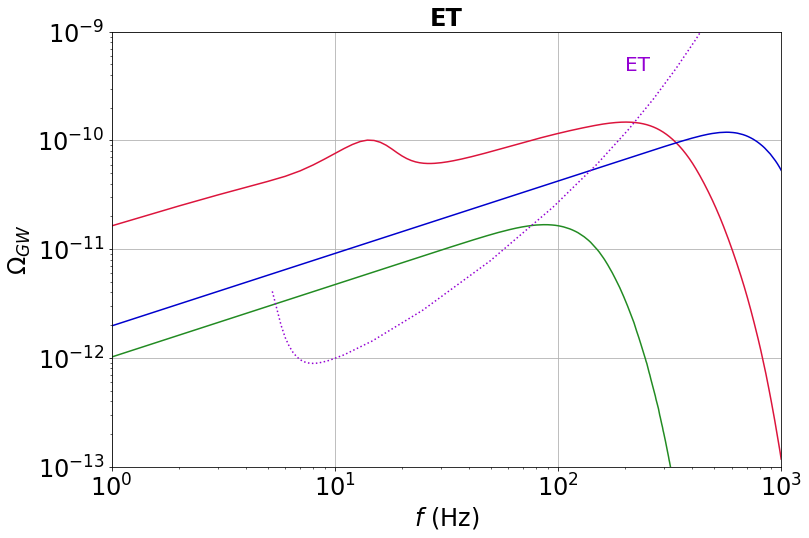}
\includegraphics[width=7.5cm]{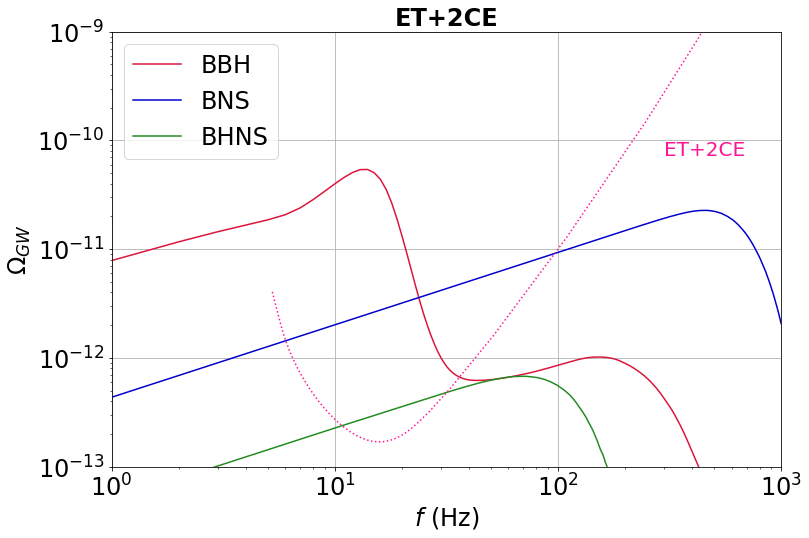}
\caption{
Energy density of ET (left panel) and ET+2CE (right panel) residual populations, at frequencies between 1Hz-2kHz. The solid lines indicate the contribution of the different types of binaries: BBH in red, BNS in blue and BHNS in green. The dotted lines show the power integrated curves for ET (in purple on the left panel) and ET+2CE (in pink on the right panel).}
\label{3G}
\end{figure}

Figure \ref{3G} shows the energy density of the residual background for ET and ET+2CE for each type of sources separately (BBHs, BNSs and BHNSs), including both populations I/II and III. For the total population, BBHs represent the larger contribution to the energy density at low frequencies, before BNSs and BHNSs. Even if they are the best detected sources, we notice that their contribution decreases but still dominates the residual background, even with third generation detectors. At 10 Hz, BBHs contribute to 95\% of the total population, 84\% of the ET residual and 94\% of the ET+2CE residual. The BNSs contribution on the other hand is 0.03\% for the total population, 10\% for the ET residual and 0.05\% for the ET+2CE residual, while the BHNS contribution is 0.03\% for the total population, 0.05\% for the ET residual and $<$0.01\% for the ET+2CE residual. 
However, because of the reduction of the background, we observe that the background from BNSs starts to exceed the BBH residual background above 400 Hz for ET and 30 Hz for ET+2CE, rather than 1000 Hz for the total background.


\subsection{Detectability}

The strategy to search for a stochastic background, which could be confounded 
with the intrinsic noise of a single interferometer, is to cross-correlate measurements of a multiple detectors.
For a network of $n$ detectors, the signal-to-noise ratio (SNR) is given by
\begin{equation}
\text{SNR} =\frac{3 H_0^2}{10 \pi^2} \sqrt{2T} \left[
\int_0^\infty df\>
\sum_{i=1}^n\sum_{j>i}
\frac{\gamma_{ij}^2(f)\Omega_{gw}^2(f)}{f^6 P_i(f)P_j(f)} \right]^{1/2}\,,
\label{eq:snrCC}
\end{equation}
In the above equations, $T$ is the observational time, $P_i$ 
and $P_j$ are the one-sided power spectral noise densities at design sensitivity of a pair of detectors $i$ and $j$ and $\gamma_{ij}$ is the normalized isotropic overlap reduction function (ORF), characterizing the loss of 
sensitivity due to the separation and the relative orientation of the detectors for sources isotropically distributed in the sky \cite{1993PhRvD..48.2389F,1992PhRvD..46.5250C}. 
Even if the cross correlation search is optimal for Gaussian backgrounds, Eq.~\ref{eq:snrCC} gives the correct expression for the background from CBCs which is not Gaussian ~\cite{2014PhRvD..89h4063M,2015PhRvD..92f3002M}.



In Table \ref{snr} we report the signal-to-noise ratio for the different residuals associated to the different networks of detectors, for an observation time of one year. We assume that we know the shape of the GW spectrum to construct the optimal filter. This assumption is realistic for population I/II for which the energy density follows a power law $\Omega_{gw} \sim f^{2/3}$ in the most sensitive frequency band, but would require accurate modelling if population III exists and is the dominant contribution.
With second generation detectors, we expect to reach an $SNR$ of about 1, in which population III contribute to less than 10\%. 

With third generation detectors, the signal-to-noise ratio increases 
due to the improvement of the sensitivity for both population I/II and III. For population I/II, it is ten times above the detection threshold of 3 $\sigma$, after a year of observation. One can notice that the SNR is a bit lower in ET+2CE ($SNR=30$) than in ET ($SNR=36$), where more sources are resolved individually.

Including the population III doubles the total SNR ($SNR=67$) in ET and improves it by a factor of about 10 in ET+2CE. The explanation is that for the high mass and high redshift population III sources, only the last stages are present in our band, making them more difficult to detect individually. However, by the time of 3G detectors, we may be using full waveforms, which will permit to increase the detectability.

In the case of LISA, with only one detector, the signal-to-noise ratio is given by \cite{Sesana:2016ljz}:
\begin{equation}
    \text{SNR} =\frac{3 H_0^2}{4 \pi^2} \sqrt{2T} \left[
\int_0^\infty df\>
\frac{\gamma^2(f)\Omega_{gw}^2(f)}{f^6 S_n^2(f)} \right]^{1/2}.
\label{eq:snrLISA}
\end{equation}
where $S_n(f)$ is the effective noise power spectral density including the sky and polarization averaged  signal  response  function \cite{Cornish:2018dyw} and $\gamma(f)$ = 1 \cite{Thrane:2013oya}.

For one year of observation are given in table \ref{snr}. We find a total signal-to-noise ratio of $SNR = 62$ for the population I/II alone and $SNR = 1588$ when we add population III.

\begin{table}
\centering\captionsetup{justification=RaggedRight, name = {Table},labelfont={bf}}
\caption{SNR values for one year of observation of each corresponding residual background.}
\begin{tabular}{|c|cc||cc||c|}
\hline
  & HLV        & HLVIK      & ET         & ET+2CE  & LISA   \\ \hline
pop I/II  &  1.02     & 1.06   & 36  & 30 & 62 \\
pop III.   & 0.08      & 0.10 &  31.2    & 255 &  1587     \\
Total & 1.08 & 1.13 & 67 & 282 & 1588 \\
\hline
\end{tabular}
\label{snr}
\end{table}

\section{Conclusion} 
In this study we have calculated the contribution of compact binary coalescences from population I/II and III, to the gravitational wave stochastic background, using the population synthesis code \textbf{\textit{StarTrack}}. 
We have used Monte Carlo techniques in order to model the evolution of the eccentricity and the redshift, and find that the eccentricity does not have a significant effect in the frequency band of ground-based detectors or even LISA ; the orbit circularizes very quickly and the higher harmonics hardly contribute very little. 
We have included the systems that do not merge during the Hubble time and find that their contribution is negligible, more than four orders of magnitude below the contribution from merging binaries.   
We obtain that the background is dominated by the population of BBHs 
and could be detected with the second generation of terrestrial detectors, HLV or HLVIK, after 7 years of observations after they have reached design sensitivity, in agreement with previous estimates \cite{implication2}. 

With future detectors, such as Einstein Telescope, Cosmic Explorer or the space antenna LISA, the goal will be to substract the background from compact binary coalescences, in order to see the cosmological or other astrophysical backgrounds below. With terrestrial detectors it will be possible to reduce the background by 2 orders of magnitude. In this case, the presence of population III systems would increase the energy density $\Omega_{GW}$ before 40 Hz by a factor of a few, changing also the shape of the spectrum between $10-40$ Hz. Modeling accurately this contribution is important to construct the optimal filter, since a simple power law like for population I/II does not apply to population III. In the LISA band, where the signal last longer, the sources overlap and it may be very difficult to separate them. However, one may be able to remove the waveform detected with terrestrial detectors at high frequencies, down to low frequencies. The challenge may be that the information on the eccentricity is lost when entering the frequency band of terrestrial detectors, but we have shown in this study that the effect of the eccentricity was small in the LISA band, giving hope the subtraction of the background would be possible.

{\em Acknowledgments}\\  
TB was supported by the TEAM/2016-3/19 grant from FNP and by the UMO-2017/26/M/ST9/00978 grant from NCN. KB acknowledges support from the Polish National Science Center (NCN) grant Maestro (2018/30/A/ST9/00050).

\bibliography{Carole}

\end{document}